\documentclass[a4paper,11pt]{article}
\pdfoutput=1 

\usepackage{jheppub} 

\usepackage[normalem]{ulem}
\usepackage[T1]{fontenc} 
\usepackage{color}

\title{No logarithmic corrections to entropy in shift-symmetric Gauss-Bonnet gravity}

\author[a]{Marek Li\v{s}ka,}
\author[b]{Robie A.~Hennigar}
\author[a]{and David Kubiz\v{n}\'{a}k}

\affiliation[a]{Institute of Theoretical Physics, Faculty of Mathematics and Physics, Charles University,
	\\V Hole\v{s}ovi\v{c}k\'{a}ch 2, 180 00 Prague 8, Czech Republic}
\affiliation[b]{Departament de F\'{i}sica Qu\'{a}ntica i Astrof\'{i}sica, Institut de Ci\'{e}ncies del Cosmos,
                Universitat de Barcelona, Mart\'{i} i Franqu\'{e}s 1, E-08028 Barcelona, Spain} 

\emailAdd{liska.mk@seznam.cz}
\emailAdd{robie.hennigar@icc.ub.edu}
\emailAdd{David.Kubiznak@matfyz.cuni.cz}

\abstract{Employing the covariant phase space formalism, we discuss black hole thermodynamics in four-dimensional scalar-tensor Einstein-Gauss-Bonnet gravity. We argue that logarithmic corrections to Wald entropy previously reported in this theory do not appear, due to the symmetry of the theory under constant shifts of the scalar field. Instead, we obtain the standard Bekenstein entropy of general relativity. Then, to satisfy the first law of black hole mechanics, the Hawking temperature must be modified. It has been proposed that such temperature modifications occur generically in scalar-tensor theories, due to different propagation speeds of gravitons and photons. We show that the temperature modifications also emerge in the Euclidean canonical ensemble approach to black hole thermodynamics. Notably, the boundary terms of the type we consider here can be considered in any scalar-tensor gravitational theories. Hence, we illustrate that adding a suitable boundary term to action may drastically affect black hole thermodynamics, changing both the entropy and the temperature.}


\begin{document}
\maketitle
	
\section{Introduction}
	
Formulations of Einstein-Gauss-Bonnet gravity in {\em four} spacetime dimensions have been studied in a number of recent papers. The original idea of rescaling the Gauss-Bonnet coupling constant as $\alpha\to\left(n-4\right)\alpha$~\cite{Glavan:2020} unfortunately does not lead to a theory derivable from a diffeomorphism invariant action. Furthermore, the theory has a number of pathologies~\cite{Ai:2020,Gurses:2020,Mahapatra:2020,Shu:2020,Arrechea:2021}. Nevertheless, its physical consequences have been extensively explored, including its black holes solutions and their thermodynamics~\cite{Wei:2020,Yerra:2022,Bravo:2022}.
	
On the one hand, a consistent purely metric, local, diffeomorphism invariant generalisation of Einstein-Gauss-Bonnet gravity in $4$ dimensions appears to be impossible. On the other hand, starting from the Einstein-Gauss-Bonnet gravity in $n>4$ dimensions and compactifying $n-4$ dimensions yields a consistent four-dimensional theory~\cite{Kobayashi:2020,Lu:2020,Hennigar:2020, Fernandes:2020nbq}. The result is special case of Horndeski gravity \cite{Horndeski:1974wa}, i.e., a local, diffeomorphism invariant, {\em scalar-tensor} theory that generalises the Einstein-Gauss-Bonnet gravity to four dimensions in a way free of basic pathologies.\footnote{Alternatively, a possibility of formulating a diffeomorphism symmetry breaking theory has been considered in \cite{Coriano:2023}; see also a recent proposal for an alternative limit of Einstein gravity to two dimensions \cite{Boehmer:2023}.} In fact, this procedure leads to a well-defined scalar-tensor Einstein-Gauss-Bonnet action in any spacetime dimension $n\ge2$. Remarkably, the static, spherically symmetric black hole solutions in four dimensions coincide with the ones found in the coupling rescaling approach~\cite{Wei:2020,Lu:2020,Hennigar:2020}. Thermodynamics of static, spherically symmetric black holes has been also reported to agree in both approaches~\cite{Wei:2020}. Notably, Wald entropy found for scalar-tensor Einstein-Gauss-Bonnet gravity contains a term logarithmic in the horizon area. Since {\em logarithmic corrections} to black hole entropy appear as the leading order quantum gravitational correction on very general grounds~\cite{Kaul:2000,Solodukhin:2011,Sen:2013,Faulkner:2013}, it would be rather interesting to recover them for a local {\em classical} modified theory of gravity.
	
Herein, elaborating on our previous short letter~\cite{MY}, we take a critical look at this reported logarithmic correction to entropy. By applying the covariant phase space formalism~\cite{Wald:1990,Wald:1993,Wald:1994,Iyer:1996}, we derive the first law of black hole mechanics and the expression for Wald entropy. We argue that the presence of logarithmic term in entropy leads to inconsistencies. The key point is that the equations of motion possess a {\em symmetry} under {\em constant shifts} of the scalar field. This symmetry is not respected by Wald entropy, unless one introduces an extra boundary term in the action. Then, Wald entropy reduces to Bekenstein entropy of general relativity. Since the total mass of the spacetime, being defined at asymptotic infinity, is also the same as the ADM mass in general relativity, fulfilling the first law of black hole mechanics requires {\em modifying} the black hole {\em temperature}\footnote{The recovery of area law in shift symmetric Horndeski theories was recently noted~\cite{Maeda:2023}. However, they do not discuss temperature modifications.}. Such modifications to temperature have been argued to occur generically in Horndeski theories, due to the propagation speed of gravitons no longer being equal to the speed of light~\cite{Liberati:2020}. Interestingly, the same modified temperature, that we fixed `add hoc' to fulfil the first law of thermodynamics, can be derived via Euclidean methods, by employing the Euclidean canonical ensemble approach \cite{BrownYork}. This provides yet another independent indication that the black hole temperature should be modified in this theory.

Although we use a particular form of 4D scalar-tensor Einstein-Gauss-Bonnet gravity as an example, our conclusions regarding the effects of shift symmetry on black hole thermodynamics in principle apply to {\em any shift symmetric} theory of gravity. In particular, this includes theories with the 4D action of the form 
\begin{equation} 	
I\propto\int d^4\!x \sqrt{-\mathfrak{g}}\left(R+\alpha\phi\mathcal{G}+\dots\right)\,,
\end{equation}
and/or to 2D Liouville gravity, 
\begin{equation}
I\propto\int d^2\!x \sqrt{-\mathfrak{g}}\Bigl(\phi R+\dots\Bigr)\,,    
\end{equation}
where $R$ denotes the Ricci scalar, $\mathcal{G}$ is the Gauss-Bonnet invariant and dots denote `arbitrary terms' that may depend on derivatives of $\phi$ but not $\phi$ itself. Moreover, these ambiguities in principle appear even in theories not possessing the shift symmetry. We will have more to say about this in the discussion.

The paper is organised as follows. First, in section~\ref{Noether}, we recall the basics of the covariant phase space formalism which offers a consistent way to derive the Smarr formula and the first law of black hole mechanics. Section~\ref{theory} reviews the formulation of scalar-tensor Einstein-Gauss-Bonnet gravity we work with. In section~\ref{EGB Noether}, we derive the covariant phase space formalism for this theory, including the Noether current and charges. Then, we apply it in section~\ref{conserved} to find expressions for mass, angular momentum and Wald entropy in stationary black hole spacetimes. We also argue why Wald entropy should coincide with Bekenstein entropy. We derive the Smarr formula and the first law of black hole mechanics for static, spherically symmetric black hole spacetimes in section~\ref{black hole}, and show that our proposal for entropy requires a modified temperature. Section~\ref{Brown-York} further supports the modified temperature proposal by obtaining it from a grandcanonical Euclidean ensemble calculation. Finally, section~\ref{discussion} sums up our findings and touches on some possible future developments.
	
Throughout the paper, we set $G=\hbar=c=k_{\text{B}}=1$, $\mu_0=1/4\pi$, and use metric signature $\left(-,+,...,+\right)$. Unless specified otherwise, we keep the spacetime dimension $n$ arbitrary. Other conventions follow reference~\cite{MTW}.

\section{Covariant phase space formalism}
\label{Noether}
	
Before going to the expressions for Einstein-Gauss-Bonnet gravity, we start by reviewing the basics of the covariant phase space formalism in general setting. Consider an $n$-dimensional manifold equipped with a volume $n$-form, $\boldsymbol{\varepsilon}$. Introduce a Lagrangian density, $L$, constructed from a collection of dynamical variables, $\phi$, and their covariant derivatives, $\nabla_{\mu}$, which satisfy $\nabla_{\mu}\boldsymbol{\varepsilon}=0$ (clearly, $\nabla_{\mu}$ is not unique). An arbitrary variation of the dynamical variables, $\delta_{1}\phi$, leads to a change in $\boldsymbol{L}$
\begin{equation}
\label{dL g}
\delta_{1}\boldsymbol{L}=A_{\phi}\delta_{1}\phi+\nabla_{\mu}\theta^{\mu}\left[\delta_{1}\right].
\end{equation}
It is easy to see that $A_{\phi}=0$ correspond to the equations of motion for $\phi$, whereas the second term $\nabla_{\mu}\theta^{\mu}\left[\delta_{1}\right]$ contributes only a boundary integral to the variation of the action. Vector density $\theta^{\mu}\left[\delta_{1}\right]$ is called the symplectic potential~\cite{Wald:1990}.
	
Next, introduce a second independent variation of the dynamical variables, $\delta_2\phi$. The commutator of the variations acting on $L$ yields
\begin{equation}
\left(\delta_1\delta_2-\delta_2\delta_1\right)L=\delta_{1}A_{\phi}\delta_{2}\phi-\delta_{2}A_{\phi}\delta_{1}\phi+\nabla_{\mu}\Omega^{\mu}\left[\delta_1,\delta_2\right],
\end{equation}
where
\begin{equation}
\label{omega g}
\Omega^{\mu}\left[\delta_{1},\delta_{2}\right]=\delta_{1}\theta^{\mu}\left[\delta_{2}\right]-\delta_{2}\theta^{\mu}\left[\delta_{1}\right],
\end{equation}
is the symplectic current~\cite{Wald:1990}. Finally, integrating $\Omega^{\mu}\left[\delta_{1},\delta_{2}\right]$ over an initial data surface $\mathcal{C}$ yields a symplectic form\footnote{The form defined in this way can be degenerate~\cite{Wald:1990}. To make it a well-defined symplectic form, we must restrict it from the space of field configurations to the space of solutions. However, this subtlety is not relevant for our purposes.}
\begin{equation}
\Omega\left[\delta_{1},\delta_{2}\right]=\int_{\mathcal{C}}\Omega^{\mu}\text{d}\mathcal{C}_{\mu}.
\end{equation}
Suppose that one of the variations is generated by a vector field $\xi^{\mu}$, i.e., $\delta_1\phi=\pounds_{\xi}\phi$. If Hamiltonian $H_{\xi}$ corresponding to evolution along $\xi^{\mu}$ exists, its change under an arbitrary variation of the dynamical variables, $\delta_{2}\phi=\delta\phi$, equals
\begin{equation}
\label{dH0}
\delta H_{\xi}=\Omega\left[\pounds_{\xi},\delta\right].
\end{equation}
by the virtue of the Hamilton equations of motion~\cite{Wald:1990}.
	
For a diffeomorphism invariant theory, any vector field $\xi^{\mu}$ with continuous first derivatives also generates a local infinitesimal symmetry of the theory~\cite{Wald:1990,Wald:1993}. Then, we can define the corresponding conserved Noether current
\begin{equation}
\label{j g}
j^{\mu}_{\xi}=\theta^{\mu}\left[\delta_{\xi}\right]-L\xi^{\mu}.
\end{equation}
This current can be written as~\cite{Wald:1993,Wald:1994}
\begin{equation}
j^{\mu}_{\xi}=-A_{\phi}\pounds_{\xi}\phi+\nabla_{\nu}Q^{\nu\mu}_{\xi},
\end{equation}
where $Q^{\nu\mu}_{\xi}$ is an antisymmetric tensor density. It follows that the divergence of $j^{\mu}_{\xi}$ vanishes on shell and $Q^{\nu\mu}_{\xi}$ is the Noether charge corresponding to the diffeomorphism generated by $\xi^{\mu}$. Now suppose one introduces a small perturbation that changes one solution of the equations of motion to another (in other words, the perturbation satisfies the linearised equations of motion, $\delta A_{\phi}=0$). Then it can be shown that a perturbation of the Hamiltonian, $\delta H_{\xi}$, is equal to an integral of the Noether charge variation $\delta Q^{\nu\mu}_{\xi}$ and the symplectic current $\theta^{\mu}\left[\delta\right]$ over the boundary $\partial\mathcal{C}$ of the initial data surface $\mathcal{C}$, i.e.,
\begin{equation}
\label{dH}
\delta H_{\xi}=\int_{\partial\mathcal{C}}\left(\delta Q^{\nu\mu}_{\xi}-2\xi^{\nu}\theta^{\mu}\left[\delta\right]\right)\text{d}\mathcal{C}_{\mu\nu}.
\end{equation}
If $\xi^{\mu}$ is a Killing vector field, the variation of the Hamiltonian vanishes identically. For a stationary black hole spacetime, equation~\eqref{dH} evaluated for the Killing vector field which defines the Killing horizon yields the first law of black hole mechanics. Using the expression for the Hawking temperature $T_{\text{H}}=\kappa/2\pi$, with $\kappa$ being the surface gravity of the horizon, one can then identify the Wald entropy of the black hole\footnote{We should note that this procedure is not fully consistent. The Hawking temperature represents a result of quantum physics in curved spacetime that must be identified in an otherwise fully classical expression.}
\begin{equation}
S_{\text{W}}=\frac{2\pi}{\kappa}\int_{\mathcal{H}}Q^{\nu\mu}_{\xi}\text{d}\mathcal{C}_{\mu\nu},
\end{equation}
where the integration is carried out over the spatial cross-section of the horizon orthogonal to the Killing vector $\xi^{\mu}$.

\section{Scalar-tensor Einstein-Gauss-Bonnet gravity in four dimensions}
\label{theory}
	
In this section, we recall the basic properties of electrovacuum scalar-tensor Einstein-Gauss-Bonnet gravity in four spacetime dimensions. The theory can be obtained by a Kaluza-Klein reduction of electrovacuum Einstein-Gauss-Bonnet gravity in any higher dimension to four dimensions (see, e.g.~\cite{Lu:2020,Hennigar:2020} for details). The resulting action reads\footnote{It is straightforward to show that dimensional reduction of the Einstein-Gauss-Bonnet-Maxwell theory does not lead to any coupling between the Maxwell term and the scalar field.}
\begin{align}
\nonumber I_{\text{EGB}}=&\frac{1}{16\pi}\int\bigg[R-2\Lambda+\alpha\phi\mathcal{G}+\alpha\Big(4G^{\lambda\rho}\nabla_{\lambda}\phi\nabla_{\rho}\phi-4\nabla_{\lambda}\phi\nabla^{\lambda}\phi\nabla^{\rho}\nabla_{\rho}\phi \\
&+2\nabla_{\lambda}\phi\nabla^{\lambda}\phi\nabla_{\rho}\phi\nabla^{\rho}\phi\Big)-F_{\lambda\rho}F^{\lambda\rho}\bigg]\sqrt{-\mathfrak{g}}\text{d}^{4}x, \label{action}
\end{align}
where
\begin{equation}
\mathcal{G}=R^2-4R_{\lambda\rho}R^{\lambda\rho}+R_{\lambda\rho\sigma\tau}R^{\lambda\rho\sigma\tau},
\end{equation}
denotes the Gauss-Bonnet term. The action~\eqref{action} possesses a shift symmetry, i.e., it is invariant under shifting the scalar field $\phi$ by any constant, as one can easily check. This symmetry of the action is not exact, but only up to a boundary term. To see this, note that the Gauss-Bonnet term in four spacetime dimensions can be written as a divergence of some vector field
\begin{equation}
\mathcal{G}=\nabla_{\mu}\mathcal{G}^{\mu}.
\end{equation}
Unfortunately, no closed, coordinate-independent expression for $\mathcal{G}^{\mu}$ is known~\cite{Padmanabhan:2011}. Nevertheless, introducing $\mathcal{G}^{\mu}$ allows us to easily show that shifting the scalar field by an arbitrary constant $C$, i.e., $\phi\to\phi+C$, changes the action as
\begin{equation}
\Delta I_{\text{EGB}}=C\int\nabla_{\mu}\mathcal{G}^{\mu}\sqrt{-\mathfrak{g}}\text{4}^{n}x.
\end{equation}
Of course, this term has no effect on the equations of motion. However, any boundary term added to the action affects the symplectic potential and through it the Noether currents and charges. In particular, we have~\cite{Wald:1994}
\begin{align}
\Delta\theta^{\mu}_{\text{EGB}}\left[\delta\right]=&\frac{\alpha}{16\pi}C\delta\mathcal{G}^{\mu}, \\
\Delta j^{\mu}_{\text{EGB},\xi}=&\frac{\alpha}{8\pi}C\nabla_{\nu}\left(\xi^{[\nu}\mathcal{G}^{\mu]}\right) \\
\Delta Q^{\nu\mu}_{\text{EGB},\xi}=&\frac{\alpha}{8\pi}C\xi^{[\nu}\mathcal{G}^{\mu]}.
\end{align}
Hence, the shift symmetry breaks down on the level of the covariant phase space formalism. To restore it, we may add a total divergence to the Lagrangian, in order to make it exactly invariant. We then have the following shift-invariant action
\begin{equation}
\label{right action}
I_{\text{inv}}=I_{\text{EGB}}-\frac{1}{16\pi}\int\nabla_{\mu}\left(\alpha\phi\mathcal{G}^{\mu}\right)\sqrt{-\mathfrak{g}}\text{d}^4x,
\end{equation}
where $I_{\text{EGB}}$ is given by equation~\eqref{action}. This modified action is only meaningful in four spacetime dimensions. In dimensions lower than four, the Gauss-Bonnet term vanishes identically and there is no divergence term breaking the symmetry, whereas for dimensions higher than four the symmetry is not present even on the level of the equations of motion.
	
Varying the action with respect to $\phi$ yields a scalar equation
\begin{align}
\nonumber &8\alpha\nabla_{\mu}\nabla_{\nu}\phi\left(-2\nabla^{\mu}\phi\nabla^{\nu}\phi-g^{\mu\nu}\nabla^{\lambda}\phi\nabla_{\lambda}\phi-\nabla^{\mu}\nabla^{\nu}\phi+g^{\mu\nu}\nabla^{\lambda}\nabla_{\lambda}\phi-G^{\mu\nu}\right)-8\alpha R^{\mu\nu}\nabla_{\mu}\phi\nabla_{\nu}\phi \\
&+\alpha\mathcal{G}=0. \label{eom scalar}
\end{align}
Notably, this equation provides a closed, covariant on-shell expression for $\mathcal{G}^{\mu}$
\begin{equation}
\label{G^a}
\mathcal{G}^{\mu}=8G^{\mu\nu}\nabla_{\nu}\phi-8\nabla^{\nu}\nabla_{\nu}\phi\nabla^{\mu}\phi+8\nabla^{\nu}\phi\nabla_{\nu}\nabla^{\mu}\phi+8\nabla_{\nu}\phi\nabla^{\nu}\phi\nabla^{\mu}\phi+\nabla_{\nu}\mathcal{F}^{\nu\mu},
\end{equation}
where arbitrary antisymmetric tensor $\mathcal{F}^{\nu\mu}$ represents an ambiguity in $\mathcal{G}^{\mu}$. We will see that this equation allows us to write the on shell Noether charges in a simple, manifestly covariant form.
	
Continuing, we vary the action with respect to the metric to obtain the following equations
\begin{align}
\nonumber & G_{\mu\nu}+\Lambda g_{\mu\nu}+2\alpha\phi\bigg[R_{\mu\lambda\rho\sigma}R_{\nu}^{\;\:\lambda\rho\sigma}-2R_{\mu\lambda\nu\rho}R^{\lambda\rho}-2R_{\mu\lambda}R_{\nu}^{\;\:\lambda}+RR_{\mu\nu}-\frac{1}{4}\Big(R^2-4R_{\lambda\rho}R^{\lambda\rho} \\
\nonumber &+R_{\lambda\rho\sigma\tau}R^{\lambda\rho\sigma\tau}\Big)g_{\mu\nu}\bigg]+2\alpha\Big(2R_{\mu\;\:\nu}^{\;\:\lambda\;\:\rho}+2R_{\mu}^{\;\:\lambda}\delta^{\rho}_{\nu}+2R_{\nu}^{\;\:\lambda}\delta^{\rho}_{\mu}-2R_{\mu\nu}g^{\lambda\rho}-2R^{\lambda\rho}g_{\mu\nu}-R\delta^{\lambda}_{\mu}\delta^{\rho}_{\nu} \\
\nonumber &+Rg_{\mu\nu}g^{\lambda\rho}\Big)\nabla_{\lambda}\nabla_{\rho}\phi+2\alpha\bigg(2R_{\mu\;\:\nu}^{\;\:\lambda\;\:\rho}+2R_{\mu}^{\;\:\lambda}\delta^{\rho}_{\nu}+2R_{\nu}^{\;\:\lambda}\delta^{\rho}_{\mu}-R_{\mu\nu}g^{\lambda\rho}-2R^{\lambda\rho}g_{\mu\nu}-R\delta^{\lambda}_{\mu}\delta^{\rho}_{\nu} \\ \nonumber 	&+\frac{1}{2}Rg_{\mu\nu}g^{\lambda\rho}\bigg)\nabla_{\lambda}\phi\nabla_{\rho}\phi+2\alpha\big(2g^{\lambda\sigma}\delta^{\rho}_{\mu}\delta^{\tau}_{\nu}-2g^{\lambda\rho}\delta^{\sigma}_{\mu}\delta^{\tau}_{\nu}+g^{\lambda\rho}g^{\sigma\tau}g_{\mu\nu}-g^{\lambda\sigma}g^{\rho\tau}g_{\mu\nu}\big)\nabla_{\lambda}\nabla_{\rho}\phi \\
\nonumber & \nabla_{\sigma}\nabla_{\tau}\phi+4\alpha\left(g^{\lambda\sigma}\delta^{\rho}_{\mu}\delta^{\tau}_{\nu}+g^{\lambda\sigma}\delta^{\tau}_{\mu}\delta^{\rho}_{\nu}-g^{\sigma\tau}\delta^{\lambda}_{\mu}\delta^{\rho}_{\nu}-g^{\lambda\sigma}g^{\rho\tau}g_{\mu\nu}\right)\nabla_{\lambda}\phi\nabla_{\rho}\phi\nabla_{\sigma}\nabla_{\tau}\phi \\
& +4\alpha\nabla^{\lambda}\phi\nabla_{\lambda}\phi\nabla_{\mu}\phi\nabla_{\nu}\phi-\alpha\nabla_{\lambda}\phi\nabla^{\lambda}\phi\nabla_{\rho}\phi\nabla^{\rho}\phi g_{\mu\nu}-2F_{\mu\lambda}F_{\nu}^{\;\:\lambda}+\frac{1}{2}F_{\lambda\rho}F^{\lambda\rho}g_{\mu\nu}=0, \label{eom metric}
\end{align}
In the following, we will write these equations as $E_{\mu\nu}=0$ to simplify our notation. Likewise, we will use $E_{\phi}=0$ for equation~\eqref{eom scalar}.
	
A particular combination of the equations of motion, $E_{\mu\nu}g^{\mu\nu}+E_{\phi}$, yields the following useful condition~\cite{Hennigar:2020}
\begin{equation}
\label{trace}
4\Lambda-R-\frac{\alpha}{2}\mathcal{G}=0.
\end{equation}
This equation is scalar and completely independent both of $\phi$ and of the electromagnetic field. Hence, it offers a simple way to check whether a particular ansatz for the metric is viable, without solving the full equations of motion. We will use this in section~\ref{black hole} to discuss black hole solutions.

\section{Covariant phase space formalism for Einstein-Gauss-Bonnet gravity}
\label{EGB Noether}
	
We now apply the covariant phase space formalism formalism reviewed in section~\ref{Noether} to the case of scalar-tensor Einstein-Gauss-Bonnet gravity. To show the significance of the shift symmetry, we first do so for action $I_{\text{EGB}}$~\eqref{action}, which is shift-symmetric only up to a boundary term. The calculations necessary to obtain the covariant phase space formalism are involved but straightforward. For the symplectic potential, we find\footnote{As an aside, upon setting $\phi=0$, the results of this section also hold for the purely metric Einstein-Gauss-Bonnet gravity in any spacetime dimension $n\ge5$.}
\begin{align}
\nonumber\theta^{\mu}_{\text{EGB}}\left[\delta\right]=&\frac{\sqrt{-\mathfrak{g}}}{16\pi}\bigg\{\big[\left(g^{\mu\nu}g^{\rho\sigma}-g^{\mu\sigma}g^{\nu\rho}\right)\left(1+2\alpha\phi R\right)+4\alpha\phi (-2R^{\rho\sigma}g^{\mu\nu}+R^{\mu\sigma}g^{\nu\rho}+R^{\nu\rho}g^{\mu\sigma} \\
\nonumber &+R^{\mu\nu\rho\sigma})\big]\nabla_{\sigma}\delta g_{\nu\rho}+4\alpha\phi\nabla^{\rho}G^{\mu\nu}\delta g_{\nu\rho}+2\alpha\big(2R^{\mu\nu\lambda\rho}+4R^{\mu\nu}g^{\lambda\rho}-2R^{\lambda\mu}g^{\nu\rho} \\
\nonumber &-2R^{\nu\rho}g^{\lambda\mu}-Rg^{\mu\nu}g^{\lambda\rho}+Rg^{\lambda\mu}g^{\nu\rho}\big)\nabla_{\lambda}\phi\delta g_{\nu\rho}+2\alpha\big(2g^{\mu\nu}g^{\lambda\rho}g^{\sigma\tau}-g^{\mu\sigma}g^{\lambda\nu}g^{\rho\tau} \\
\nonumber &-g^{\nu\rho}g^{\lambda\mu}g^{\sigma\tau}-g^{\mu\nu}g^{\rho\sigma}g^{\lambda\tau}+g^{\mu\sigma}g^{\nu\rho}g^{\lambda\tau}\big)\nabla_{\lambda}\phi\nabla_{\tau}\phi\nabla_{\sigma}\delta g_{\nu\rho}+2\alpha\big(2g^{\mu\nu}g^{\lambda\sigma}g^{\rho\tau} \\
\nonumber &-2g^{\lambda\mu}g^{\nu\sigma}g^{\rho\tau}-g^{\mu\tau}g^{\nu\rho}g^{\lambda\sigma}+g^{\lambda\mu}g^{\nu\rho}g^{\sigma\tau}\big)\nabla_{\lambda}\phi\nabla_{\sigma}\nabla_{\tau}\phi\delta g_{\nu\rho}+2\alpha\big(2g^{\mu\nu}g^{\rho\sigma} \\
\nonumber &-g^{\mu\sigma}g^{\nu\rho}\big) \nabla_{\lambda}\phi\nabla_{\sigma}\phi\nabla^{\sigma}\phi\delta g_{\nu\rho}+4\alpha\big(2G^{\mu\nu}\nabla_{\nu}\phi\delta\phi-\nabla^{\nu}\phi\nabla_{\nu}\phi\nabla^{\mu}\delta\phi-2\nabla^{\nu}\nabla_{\nu}\phi\nabla^{\mu}\phi\delta\phi \\
&+2\nabla^{\nu}\phi\nabla_{\nu}\nabla^{\mu}\phi\delta\phi+2\nabla_{\nu}\phi\nabla^{\nu}\phi\nabla^{\mu}\phi\delta\phi\big)-2F^{\mu\nu}\delta A_{\nu}\bigg\}. \label{theta}
\end{align}
	
Next, we want to find the Noether current corresponding to infinitesimal diffeomorphism transformations. However, a subtle problem occurs for fields with gauge freedom, such as $A_{\mu}$. The usual definition of an infinitesimal diffeomorphism transformation, i.e., a Lie derivative along the diffeomorphism generator $\xi^{\mu}$, is not gauge invariant. The issue can be systematically dealt with by generalising the Noether charge formalism to vector bundles~\cite{Prabhu:2017}. However, a simpler solution sufficient for our purposes lies in modifying the action of infinitesimal diffeomorphisms on $A_{\mu}$ to make it gauge invariant~\cite{Elgood:2020}. We have
\begin{equation}
\delta_{\xi}A_{\mu}=\pounds_{\xi}A_{\mu}-\nabla_{\mu}\left(A_{\nu}\xi^{\nu}-P^{\text{EM},\xi}\right),
\end{equation}
where $P^{\text{EM},\xi}$, is a gauge invariant quantity satisfying~\cite{Elgood:2020}
\begin{equation}
k^{\nu}F_{\nu\mu}=-\nabla_{\mu}P^{\text{EM},k},
\end{equation}
for any vector field $k^{\mu}$ generating a symmetry of all the dynamical fields. For a suitable gauge fixing of potentials, we have simply $P^{\text{EM},\xi}=A_{\lambda}\xi^{\lambda}$, and the gauge invariant diffeomorphism transformation reduces to the standard Lie derivative, but this is not generally the case.
	
We are now ready to compute the Noether current corresponding to an infinitesimal diffeomorphism transformation generated by a vector field $\xi^{\mu}$. Starting from the general definition~\eqref{j g}, we get
\begin{align}
\nonumber j^{\mu}_{\text{EGB},\xi}=&\frac{\sqrt{-\mathfrak{g}}}{8\pi}\left(\xi^{\nu}E_{\nu}^{\;\:\mu}-2P^{\text{EM},\xi}\nabla_{\nu}F^{\nu\mu}\right)+\frac{\sqrt{-\mathfrak{g}}}{16\pi}\nabla_{\nu}\bigg\{\nabla^{[\nu}\xi^{\mu]}\left(2+4\alpha\phi R\right)-16\alpha\phi R^{\lambda[\nu}\nabla_{\lambda}\xi^{\mu]} \\
\nonumber &+4\alpha\phi R^{\nu\mu\lambda\rho}\nabla_{\lambda}\xi_{\rho}-8\alpha R\nabla^{[\nu}\phi\xi^{\mu]}+16\alpha\nabla_{\lambda}\phi R^{\lambda[\nu}\xi^{\mu]}+16\alpha R_{\lambda}^{\;\:[\mu}\nabla^{\nu]}\phi\xi^{\lambda} \\
\nonumber &+8\alpha R^{\mu\nu\lambda\rho}\nabla_{\lambda}\phi\xi_{\rho}+4\alpha\nabla^{[\mu}\xi^{\nu]}\nabla_{\lambda}\phi\nabla^{\lambda}\phi+8\alpha\nabla_{\lambda}\xi^{[\mu}\nabla^{\nu]}\phi\nabla^{\lambda}\phi+8\alpha\xi^{[\mu}\nabla^{\nu]}\nabla^{\lambda}\phi\nabla_{\lambda}\phi \\
&-8\alpha\xi^{[\mu}\nabla^{\nu]}\phi\nabla^{\lambda}\nabla_{\lambda}\phi+8\alpha\xi^{\lambda}\nabla_{\lambda}\nabla^{[\mu}\phi\nabla^{\nu]}\phi+8\alpha\nabla_{\lambda}\phi\nabla^{\lambda}\phi\nabla^{[\nu}\phi\xi^{\mu]}+2F^{\nu\mu}P^{\text{EM},\xi}\bigg\}. \label{j EGB}
\end{align}
The first two terms are proportional to the equations of motion for the metric and the electromagnetic field. Note that the equations of motion for the scalar field $\phi$ as well as the second set of the Maxwell equations $\nabla_{[\mu}F_{\nu\rho]}=0$ do not appear in the Noether current.
	
The total divergence term in the Noether current~\eqref{j EGB} gives us directly the Noether charge antisymmetric tensor density
\begin{align}
\nonumber Q^{\nu\mu}_{\text{EGB},\xi}=&\frac{\sqrt{-\mathfrak{g}}}{16\pi}\bigg[\nabla^{[\nu}\xi^{\mu]}\left(2+4\alpha\phi R\right)-16\alpha\phi R^{\lambda[\nu}\nabla_{\lambda}\xi^{\mu]}+4\alpha\phi R^{\nu\mu\lambda\rho}\nabla_{\lambda}\xi_{\rho}-8\alpha R\nabla^{[\nu}\phi\xi^{\mu]} \\
\nonumber &+16\alpha\nabla_{\lambda}\phi R^{\lambda[\nu}\xi^{\mu]}+16\alpha R_{\lambda}^{\;\:[\mu}\nabla^{\nu]}\phi\xi^{\lambda}+8\alpha R^{\mu\nu\lambda\rho}\nabla_{\lambda}\phi\xi_{\rho}+4\alpha\nabla^{[\mu}\xi^{\nu]}\nabla_{\lambda}\phi\nabla^{\lambda}\phi \\
\nonumber &+8\alpha\nabla_{\lambda}\xi^{[\mu}\nabla^{\nu]}\phi\nabla^{\lambda}\phi+8\alpha\xi^{[\mu}\nabla^{\nu]}\nabla^{\lambda}\phi\nabla_{\lambda}\phi-8\alpha\xi^{[\mu}\nabla^{\nu]}\phi\nabla^{\lambda}\nabla_{\lambda}\phi \\
&+8\alpha\xi^{\lambda}\nabla_{\lambda}\nabla^{[\mu}\phi\nabla^{\nu]}\phi+8\alpha\nabla_{\lambda}\phi\nabla^{\lambda}\phi\nabla^{[\nu}\phi\xi^{\mu]}+2F^{\nu\mu}P^{\text{EM},\xi}\bigg].
\end{align}
	
In the previous section, we have shown that the symplectic potential $\theta^{\mu}_{\text{EGB}}\left[\delta\right]$, the Noether current $j^{\mu}_{\text{EGB},\xi}$, and the Noether charge $Q^{\nu\mu}_{\text{EGB},\xi}$ are not invariant under shifting the scalar field by a constant, $\phi\to\phi+C$. Since the equations of motion and, therefore, all the solutions of the theory do not change under such a shift, it is somewhat unsatisfactory that the locally defined Noether currents and charges, which are used to define measurable quantities such as mass and entropy, do not share this invariance. The solution lies in deriving the symplectic potential from the action $I_{\text{inv}}$~\eqref{right action}, which is exactly shift-invariant. Then, we obtain
\begin{align}
\theta^{\mu}_{\text{inv}}\left[\delta\right]=&\theta^{\mu}_{\text{EGB}}\left[\delta\right]-\frac{\alpha}{16\pi}\phi\delta\mathcal{G}^{\mu}, \label{right theta}\\
j^{\mu}_{\text{inv},\xi}=&j^{\mu}_{\text{EGB},\xi}-\frac{\alpha}{8\pi}\nabla_{\nu}\left(\phi\xi^{[\nu}\mathcal{G}^{\mu]}\right) \\
Q^{\nu\mu}_{\text{inv},\xi}=&Q^{\nu\mu}_{\text{EGB},\xi}-\frac{\alpha}{8\pi}\phi\xi^{[\nu}\mathcal{G}^{\mu]}, \label{right charge}
\end{align}
where we recall that $\mathcal{G}^{\mu}$ is defined so that its divergence equals the Gauss-Bonnet term, $\mathcal{G}=\nabla_{\mu}\mathcal{G}^{\mu}$. The expressions $\theta^{\mu}_{\text{inv}}\left[\delta\right]$, $j^{\mu}_{\text{inv},\xi}$ and $Q^{\nu\mu}_{\text{inv},\xi}$ have the advantage of being explicitly invariant under constant shifts of the scalar field. On shell, using equation~\eqref{G^a}, we can express $\mathcal{G}^{\mu}$ in terms of derivatives of $\phi$ and curvature tensors, making the formulas for the symplectic potential and Noether currents and charges explicitly covariant.
		
Of course, the covariant phase space formalism is also affected by other boundary contributions to the action, most notably by the Gibbons-Hawking-York boundary term. For the scalar-tensor Einstein-Gauss-Bonnet theory we consider, this term has been derived in the literature~\cite{Ma:2020}. In section~\ref{black hole}, we show that it has no effect on thermodynamics of static, spherically symmetric black holes. We expect this result to hold in general as well, since it requires only regularity of the Gibbons-Hawking-York term on the horizon and its sufficiently fast fall-off in the asymptotic region~\cite{Khodabakhshi:2020fhb}.
	
One might wonder whether the shift symmetry also carries a nontrivial Noether charge that might affect the covariant phase space formalism. Starting from the symplectic potential~\eqref{right theta}, and setting $\delta_{C}\phi=C$, $\delta_{C}g_{\mu\nu}=0$, we easily obtain the Noether current corresponding to shift symmetry
\begin{equation}
j^{\mu}_{C}=\alpha C\left(8G^{\mu\nu}\nabla_{\nu}\phi-8\nabla^{\nu}\nabla_{\nu}\phi\nabla^{\mu}\phi+8\nabla^{\nu}\phi\nabla_{\nu}\nabla^{\mu}\phi+8\nabla_{\nu}\phi\nabla^{\nu}\phi\nabla^{\mu}\phi-\alpha\mathcal{G}^{\mu}\right).
\end{equation}
On shell, the Noether current vanishes due to equation~\eqref{G^a}. Hence, there is no nontrivial Noether charge associated with shift symmetry and it will not generate any contributions to conserved quantities.
	
At this point, we have a complete covariant phase space formalism for electrovacuum scalar-tensor Einstein-Gauss-Bonnet theory. Deriving an explicit expression for the symplectic current would again be straightforward (albeit laborious), but we do not need it for our purposes. We stress that the expression for the symplectic form is not affected by the boundary term we added to the action -- it is explicitly shift symmetric regardless whether we derive it from $\theta^{\mu}_{\text{EGB}}\left[\delta\right]$ or $\theta^{\mu}_{\text{inv}}\left[\delta\right]$. The difference in both expressions is proportional to $\left(\delta_{1}\delta_{2}-\delta_{2}\delta_{1}\right)\mathcal{G}^{\mu}$, see equation~\eqref{omega g}, which vanishes identically. Therefore, the fundamental theorem of covariant phase space formalism~\cite{Compere:2018}, which asserts the uniqueness of the symplectic form, is respected. In the remainder of the paper, we apply this formalism to the particular case of black hole spacetimes. The general formalism can be, of course, equally applied to any other setting of interest, such as spacetime causal diamonds, cosmological spacetimes, or gravitational waves.

\section{Conserved quantities in stationary black hole spacetimes}
\label{conserved}
	
\subsection{Mass and angular momentum}	
	
We are ready to identify the conserved quantities in 4D scalar-tensor Einstein-Gauss-Bonnet gravity. We will do so on the example of an asymptotically flat, stationary spacetime containing a single black hole. In particular, we require that there exist a coordinate system in which one can write the metric as $g_{\mu\nu}=\eta_{\mu\nu}+O\left(1/r\right)$. Consequently, the Riemann tensor behaves asymptotically as $O\left(1/r^{3}\right)$. Inspecting equations of motion for the metric~\eqref{eom metric} and the scalar field~\eqref{eom scalar} we can see that this allows asymptotic behaviour of the scalar field at most $\phi=O\left(r^{1/4}\right)$. Otherwise, $\phi$ would source a contribution to the Riemann tensor larger than $O\left(1/r^{3}\right)$. In an asymptotically flat, stationary spacetime, the canonical mass $\mathcal{M}$ is defined as the contribution to the Hamiltonian corresponding to the evolution along the time translation Killing vector $t^{\mu}$ coming from the asymptotic infinity~\cite{Wald:1993,Wald:1994}. Therefore, we have for a small on-shell perturbation of $\mathcal{M}$ (the perturbation must be such that it does not spoil the asymptotic flatness)
\begin{equation}
\delta\mathcal{M}=\int_{\infty}\left(\delta Q^{\nu\mu}_{t}-2t^{\nu}\theta^{\mu}\left[\delta\right]\right)\text{d}\mathcal{C}_{\mu\nu}.
\end{equation}
Interestingly, all the correction terms yield vanishing contributions. Therefore, we recover the GR result for canonical mass which is equivalent to the ADM mass. In total, it holds $\mathcal{M}=M_{\text{ADM,GR}}$. The same applies to the canonical angular momentum $\mathcal{J}$, defined with respect to the rotational Killing vector $\varphi^{\mu}$
\begin{equation}
\delta\mathcal{J}=-\int_{\infty}\delta Q^{\nu\mu}_{t}\text{d}\mathcal{C}_{\mu\nu}=\delta J_{\text{ADM,GR}}.
\end{equation}
To sum up, the conserved charges of asymptotically flat spacetimes in the scalar-tensor Einstein-Gauss-Bonnet gravity defined at the asymptotic infinity are the same as their counterparts in general relativity. This is natural, since one does not expect the higher order curvature corrections to significantly affect the weak gravity regime. For the same reason, these results hold regardless of whether we consider the shift-symmetric Noether charge $Q^{\nu\mu}_{\text{inv},\xi}$ or the Noether charge $Q^{\nu\mu}_{\text{EGB},\xi}$, which changes under constant shifts of the scalar filed.

\subsection{Entropy prescription}
\label{ent section}

The situation with black hole entropy is more interesting. The definition of Wald entropy of a stationary asymptotically flat black hole spacetime reads
\begin{equation}
S_{\text{W}}=\frac{2\pi}{\kappa}\int_{\mathcal{H}}Q^{\nu\mu}_{\xi}\text{d}\mathcal{C}_{\mu\nu},
\end{equation}
where $\xi^{\mu}=t^{\mu}+\Omega_{\text{H}}\varphi^{\mu}$ is the Killing vector tangent to the horizon ($\Omega_{\text{H}}$ denotes the constant angular frequency of the horizon), $\kappa$ the corresponding surface gravity, and $\mathcal{H}$ a spatial cross-section of the horizon. 
	
We first look at the entropy computed from the Noether charge $Q^{\nu\mu}_{\text{EGB},\xi}$, which is not shift-symmetric. Using that $\nabla_{[\nu}\xi_{\mu]}=\kappa\epsilon_{\nu\mu}$, where $\epsilon_{\nu\mu}$ denotes the binormal to the horizon, we easily obtain, in agreement with the previously reported results~\cite{Wei:2020}
\begin{align}
\nonumber S_{W,\text{EGB}}=&\frac{1}{16}\int_{\mathcal{H}}\left[\epsilon^{\mu\nu}\left(2+4\alpha\phi R\right)-16\alpha\epsilon^{\lambda\mu}\phi R_{\lambda}^{\;\:\nu}+4\alpha\epsilon_{\lambda\rho}\phi R^{\nu\mu\lambda\rho}\right]\epsilon_{\nu\mu}\text{d}^2\mathcal{A} \\
=&\frac{\mathcal{A}}{4}+\frac{1}{4}\int_{\mathcal{H}}\alpha\phi\left(\epsilon^{\mu\nu}R-4\epsilon^{\lambda\mu} R_{\lambda}^{\;\:\nu}+\epsilon_{\lambda\rho}R^{\nu\mu\lambda\rho}\right)\epsilon_{\nu\mu}\text{d}^2\mathcal{A}. \label{entropy}
\end{align}
where $\text{d}^2\mathcal{A}$ is the area element on the horizon. The first term is the well known Bekenstein entropy, the second one is a Gauss-Bonnet correction. Notably, the correction term is independent of the derivatives of $\phi$. In fact, it is simply the correction to entropy obtained in the purely metric Einstein-Gauss-Bonnet gravity in dimensions higher than four, multiplied by the scalar field $\phi$. Since, as we discussed, $\phi$ can be shifted by any constant $C$, we have a constant ambiguity in the total entropy of the form
\begin{align}
\Delta S=\frac{1}{4}C\int_{\mathcal{H}}\alpha\left(\epsilon^{\mu\nu}R-4\epsilon^{\lambda\mu} R_{\lambda}^{\;\:\nu}+\epsilon_{\lambda\rho}R^{\nu\mu\lambda\rho}\right)\epsilon_{\mu\nu}\text{d}^2\mathcal{A}=4\pi\alpha C,
\end{align}
where we used that the integral is proportional to the Euler characteristic of the horizon spatial cross-section.
	
Next, we look at the entropy given by the shift-symmetric Noether charge $Q^{\nu\mu}_{\text{inv},\xi}$. Since $\xi^{\mu}$ is a null vector tangent to the horizon (and, if the horizon possesses a bifurcation surface, $\xi^{\mu}$ vanishes there), the terms in $Q^{\nu\mu}_{\xi}$ proportional to $\xi^{\mu}$ (and not containing its derivatives) should integrate to $0$ over the horizon and not contribute to Wald entropy. The difference between $Q^{\nu\mu}_{\text{inv},\xi}$ and $Q^{\nu\mu}_{\text{EGB},\xi}$ equals $-\alpha\phi\xi^{[\nu}\mathcal{G}^{\mu]}/\left(8\pi\right)$, which is proportional to $\xi^{\mu}$. Hence, it would seem that the black hole entropy does not depend on whether we work with $Q^{\nu\mu}_{\text{inv},\xi}$ or $Q^{\nu\mu}_{\text{EGB},\xi}$. However, the properties of the Gauss-Bonnet term change this picture. It has been shown that adding the boundary term $\nabla_{\mu}\mathcal{G}^{\mu}$ to the Einstein-Hilbert action in four dimensions, while leaving the Einstein equations unchanged, shifts Wald entropy of a black hole by a constant proportional to the Euler characteristic of the horizon~\cite{Jacobson:1993,Sarkar:2011}. In our case, with the Gauss-Bonnet term being multiplied by the scalar field, the entropy changes more substantively. The easiest way to see that the Gauss-Bonnet term contributes to Wald entropy is by considering its form in a spacetime with Killing symmetries. There, the vector $\mathcal{G}^{\mu}$ reads~\cite{Padmanabhan:2011}
\begin{equation}
\mathcal{G}^{\mu}=-2\frac{\partial\mathcal{G}}{\partial R_{\mu\nu\rho\sigma}}\xi_{\nu}\nabla_{\rho}\xi_{\sigma}.
\end{equation}
Hence, $\mathcal{G}^{\mu}$ depends on the derivative of the Killing vector and can contribute to Wald entropy.\footnote{Another way to see this is by noting that $\mathcal{G}^{\mu}$ diverges on the bifurcation surface of the horizon (if one exists), compensating the vanishing of $\xi^{\mu}$. The overall contribution of the Gauss-Bonnet boundary term is then finite.} The contribution to Wald entropy given by the extra term $-\alpha\phi\xi^{[\nu}\mathcal{G}^{\mu]}/\left(8\pi\right)$ in the shift-invariant Noether charge $Q^{\nu\mu}_{\text{inv},\xi}$ equals
\begin{equation}
-\frac{2\pi}{\kappa}\int_{\mathcal{H}}\alpha\phi\xi^{[\nu}\mathcal{G}^{\mu]}/\left(8\pi\right)\epsilon_{\nu\mu}\text{d}^2\mathcal{A}=-\frac{1}{4}\int_{\mathcal{H}}\alpha\phi\left(\epsilon^{\mu\nu}R-4\epsilon^{\lambda\mu} R_{\lambda}^{\;\:\nu}+\epsilon_{\lambda\rho}R^{\nu\mu\lambda\rho}\right)\epsilon_{\nu\mu}\text{d}^2\mathcal{A}.
\end{equation}
Therefore, the total Wald entropy corresponding to $Q^{\nu\mu}_{\text{inv},\xi}$ reduces to the well-known Bekenstein entropy of general relativity
\begin{equation}
\label{entropy right}
S_{\text{W,inv}}=\frac{\mathcal{A}}{4}.
\end{equation}
without any correction terms. At the first glance, this is a surprising result that differs from the ones arrived at in various studies of black hole thermodynamics in four-dimensional Einstein-Gauss-Bonnet gravity. However, this is also essentially the only entropy prescription compatible with the symmetry of the equations of motion under $\phi\to\phi+C$. Stated differently, the equations of motion imply that physics depends only on derivatives of $\phi$ and not on the value of the field itself. Hence, $\phi$ is not a physical observable. It would then be fairly strange if entropy depended on the value of $\phi$ in any way. We stress that the ambiguity one obtains when entropy depends on the value of $\phi$ on the horizon is more serious than that from adding the Gauss-Bonnet term to the Einstein-Hilbert action. In that case, the entropy is shifted by a universal constant depending only on the value of the Gauss-Bonnet coupling. However, the dependence on the arbitrary value of $\phi$ allows for two otherwise physically identical black hole spacetimes to have different entropies. Moreover, we are free to choose a suitable negative $\phi$ on the horizon to make the overall entropy of the black hole zero (or even negative). Since the entropy of the (minimally coupled) matter outside of the black hole does not depend on the value of $\phi$, a small variation of $\phi$ can in principle violate the second law of thermodynamics. An alternative argument for the area law entropy in shift symmetric theories was recently proposed~\cite{Maeda:2023}. It suggests that, since the value of the scalar field is arbitrary, it is natural to eliminate its contribution to entropy by setting it to zero on the horizon.

Since we do not have a general, manifestly covariant expression for $\mathcal{G}^{\mu}$, one might in principle worry whether the entropy formula we obtained respects the covariance. However, Wald entropy is an on shell quantity. Then, the on shell expression for $\mathcal{G}^{\mu}$~\eqref{G^a} guarantees that entropy (and, hence, the first law of black hole mechanics) remains fully covariant. Equation~\eqref{G^a} does contain an ambiguity given by divergence of an antisymmetric tensor $\mathcal{F}^{\nu\mu}$, but it can be shown that this ambiguity does not affect the conserved quantities. To see this, note that the ambiguity in the conserved quantities (both at the Killing horizon and at the asymptotic infinity) would be proportional to $\xi^{[\mu\vert}\partial_{\lambda}\left(\sqrt{-\mathfrak{g}}\mathcal{F}^{\lambda\vert\nu]}\right)\xi_{[\mu}n_{\nu]}$, where $n^{\mu}$ denotes a spacelike vector normal to the boundary and we used that covariant derivative of an antisymmetric tensor density of weight $1$ reduces to a partial derivative. Any contributions to $\mathcal{F}^{\lambda\nu}$ orthogonal to bi-normal $\xi_{[\mu}n_{\nu]}$ contract to zero. Hence, we can only study $\mathcal{F}^{\lambda\nu}=f\xi^{[\lambda}n^{\nu]}$, where $f$ is an arbitrary function. If we adapt the coordinate system so that $\partial_{\lambda}n^{\nu}=\partial_{\lambda}\xi^{\nu}=0$, we are left with contribution $\xi^{\mu}\xi^{[\lambda}n^{\nu]}\partial_{\lambda}\left(\sqrt{-\mathfrak{g}}f\right)\xi_{[\mu}n_{\nu]}=\xi^{\mu}n^{\nu}\xi^{\lambda}\partial_{\lambda}\left(\sqrt{-\mathfrak{g}}f\right)\xi_{[\mu}n_{\nu]}$. This is proportional to the Lie derivative of $\sqrt{-\mathfrak{g}}f$ with respet to Killing vector $\xi^{\lambda}$. If the Killing symmetry is respected, this Lie derivative must vanish. Therefore, the ambiguity in $\mathcal{G}^{\mu}$ does not affect any of the conserved quantities.

To sum up, the consistency of black hole thermodynamics in scalar-tensor Einstein-Gauss-Bonnet gravity seems to favour the unmodified entropy prescription~\eqref{entropy right}. In the following, we study the implications of this choice of entropy on the example of a static, spherically symmetric black hole spacetime. We will see that it requires a modified prescription for the black hole temperature.

\section{Spherical black holes}	\label{black hole}
	
\subsection{Black hole solutions}	
	
We now apply the general covariant phase space formalism to the particular case of static, spherically symmetric, electrovacuum black hole solutions in four spacetime dimensions. For the sake of clarity, we consider asymptotically flat spacetimes. Our treatment can be straightforwardly generalised to the asymptotically anti-de Sitter case, upon introducing a suitable regularisation~\cite{Kubiznak:2015} (the asymptotically de Sitter case must be treated separately due to the presence of the cosmological horizon).
	
Any static, spherically symmetric metric can be written as
\begin{equation}
\text{d}s^2=-f\left(r\right)\text{d}t^2+\frac{\text{d}r^2}{f\left(r\right)h\left(r\right)}+r^2\left(\text{d}\theta^2+\sin^2\theta\text{d}\phi^2\right).
\end{equation}
where $f\left(r\right)$, $h\left(r\right)$ are arbitrary functions. Here we are interested only in the solutions with $h\left(r\right)=1$, which can be found analytically.
	
The requirements of staticity and spherical symmetry together with the Maxwell equations imply $F_{tr}=-F_{rt}=Q/r^2\sqrt{h}$, where $Q$ is the constant electric charge and, for $h=1$, we have $P^{\text{EM},\xi}=-Q/r$. We note that this result does not depend on the equations of motion for the metric~\eqref{eom metric} and the scalar field~\eqref{eom scalar}, and applies to static, spherically symmetric solutions of any gravitational theory, provided that the electromagnetic field is minimally coupled.
	
To determine the function $f$, we may first use condition~\eqref{trace} which is independent of the scalar field $\phi$. It implies
\begin{equation}
\alpha f^2-\left(r^2+2\alpha\right)f-C_2-C_1r+r^2=0,
\end{equation}
which has a general solution
\begin{equation}
\label{f general}
f_{\pm}=1+\frac{r^2}{2\alpha}\left(1\pm\sqrt{1+\frac{4\alpha C_1}{r^3}+\frac{4\alpha\left(C_2+\alpha\right)}{r^4}}\right).
\end{equation}
Plugging this ansatz into the equations of motion\footnote{In practice, it is easier to first use the symmetries to simplify the action and then derive the equations of motion specialised for the static, spherically symmetric case~\cite{Lu:2020,Hennigar:2020}.} yields three independent conditions. Two of them are solved by setting $\phi$ to~\cite{Lu:2020,Hennigar:2020}
\begin{equation}
\label{phi}
\phi\left(r\right)=\ln\left(\frac{r}{L}\right)\pm\int_{r_+}^{r}\frac{1}{\rho\sqrt{f\left(\rho\right)}}\text{d}\rho.
\end{equation}
The third equation yields a further constrain on $f$
\begin{equation}
\left(r^3+2\alpha r-2\alpha rf\right)f'+\left(r^2-2\alpha+\alpha f\right)f+\alpha-r^2+Q^2=0,
\end{equation}
where $'$ denotes a partial derivative with respect to $r$. The ansatz~\eqref{f general} for $f$ solves this equation provided we set $C_2=-Q^2-\alpha$. In the following we study only the asymptotically anti-de Sitter/flat branch of the solution which corresponds to the minus sign between the terms in the brackets in ansatz~\eqref{f general}. This branch approaches the Schwarzschild-anti-de Sitter black hole in the limit of $\alpha\to0$. The last unspecified constant is then related to Schwarzschild mass parameter, $C_1=2M$. As we have shown in the previous section, the Schwarzschild mass parameter also gives the canonical mass of the black hole solution we consider. In total, we have the following solution for $f$:
\begin{equation}
\label{f solution}
f=1+\frac{r^2}{2\alpha}\left(1-\sqrt{1+\frac{8\alpha M}{r^3}-\frac{4\alpha Q^2}{r^4}}\right).
\end{equation}
By computing the Kretschmann scalar, one can easily check that even with the Gauss--Bonnet corrections, the resulting black hole spacetime remains singular in the origin at $r=0$. Function $f$ has two real roots. The larger one corresponds to the location of the event horizon, the smaller to the inner, Cauchy horizon. In between the horizons, $f$ is negative. The event horizon is a Killing horizon with respect to the time translational Killing vector $t^{\mu}=\left(1,0,0,0\right)$.

\subsection{Thermodynamics and modified temperature}	
	
Let us now look at the thermodynamics of these black hole solutions. Integrating the on-shell identity $j^{\mu}_{\text{inv},t}=\nabla_{\nu}Q^{\nu\mu}_{\text{inv},t}$ over a spatial Cauchy surface $\mathcal{C}$ connecting the spatial infinity and the black hole event horizon yields the Smarr formula\footnote{If we instead use the $\phi$-dependent Noether current $j^{\mu}_{\text{EGB},t}$ and charge $Q^{\nu\mu}_{\text{EGB},t}$, the integral of $Q^{\nu\mu}_{\text{EGB},t}$ over the horizon contains an extra term $4\pi\alpha\ln\frac{r_{+}}{L}$. The same extra term appears in the volume integral of $j^{\mu}_{\text{EGB},t}$. Hence, the Smarr formula we find reads
\begin{equation}
\label{smarr}
\frac{M}{2}-\frac{\kappa}{2\pi}\left(\frac{\mathcal{A}}{4}+4\pi\alpha\ln\frac{r_{+}}{L}\right)-\frac{Q^2}{2r_{+}}=\frac{\alpha}{2r_{+}}-2\alpha\kappa\ln\frac{r_{+}}{L}+\alpha\kappa,
\end{equation}
which agrees with the previously reported result~\cite{Wei:2020,Yerra:2022}. The only difference is that the other references chose to set $L=\sqrt{\alpha}$, although the constant $L$ is in fact completely arbitrary.}
\begin{align}
\int_{\infty}Q^{\nu\mu}_{\text{inv},t}\text{d}\mathcal{C}_{\mu\nu}-\int_{\mathcal{H}}Q^{\nu\mu}_{\text{inv},t}\text{d}\mathcal{C}_{\mu\nu}&=\int_{\mathcal{C}}j^{\mu}_{\text{inv},t}\text{d}\mathcal{C}_{\mu} \\
\frac{M}{2}-\frac{\kappa}{2\pi}\frac{\mathcal{A}}{4}+\frac{Q^2}{2r_{+}}&=\frac{Q^2}{r_{+}}+\alpha\kappa+\frac{\alpha}{2r_{+}}.
\end{align}
where the non-zero contribution to the volume integral of $j^{\mu}_{\text{inv},t}=\theta^{\mu}_{\text{inv}}\left[\delta_{t}\right]-L_{\text{inv}}t^{\mu}$ comes from the Lagrangian $L_{\text{inv}}$ evaluated for the black hole solution we study.
	
The first law of black hole mechanics can be derived from the equation~\eqref{dH} for variation $\delta H_{t}$ of the Hamiltonian corresponding to evolution along Killing vector $t^{\mu}$
\begin{align}
\int_{\infty}\left(\delta Q^{\nu\mu}_{t}-2t^{\nu}\theta^{\mu}\left[\delta\right]\right)\text{d}\mathcal{C}_{\mu\nu}-\int_{\mathcal{H}}\left(\delta Q^{\nu\mu}_{t}-2t^{\nu}\theta^{\mu}\left[\delta\right]\right)\text{d}\mathcal{C}_{\mu\nu}=&0, \\
\delta M-\frac{\kappa}{2\pi}\left(1+\frac{2\alpha}{r_+^2}\right)\frac{\delta\mathcal{A}}{4}-\frac{Q}{r_+}\delta Q=&0. \label{first law}
\end{align}
In deriving this equation we assume, as is usually done~\cite{Bardeen:1973,Wald:1993}, that the surface gravity $\kappa$ is held fixed, i.e., $\delta\kappa=0$. We would like to interpret the first law of black hole mechanics~\eqref{first law} as the first law of black hole thermodynamics. We can easily identify the last term as the electric field work contribution, $\Phi\delta Q$, where $\Phi=Q/r_+$ denotes the electric potential on the horizon. The mass term $\delta M$ corresponds to a variation of the black hole enthalpy~\cite{Kubiznak:2015} (though no variation of $\Lambda$ is considered here). The second term should then be interpreted as $T_{\text{H}}\delta S_{\text{W,EGB}}$, where $T_{\text{H}}$ denotes the Hawking temperature at which the black hole radiates. The standard calculations of the black hole radiation yield for the Hawking temperature $\kappa/\left(2\pi\right)$. However, our form of the first law~\eqref{first law} implies
\begin{equation}
\label{T mod}
T_{\text{H}}=\frac{\kappa}{2\pi}\left(1+\frac{2\alpha}{r_+^2}\right).
\end{equation}
As the Hawking temperature should depend on the kinematic features of the black hole spacetime and not on the gravitational dynamics~\cite{Visser:2003}, the idea of modifying it in this case may seem curious. However, it is well known that scalar-tensor theories of gravity generically lead to propagation speeds of gravitons which differ from the speed of light. It has been pointed out that, since gravitons contribute to black hole radiation, the change in their propagation speed can affect the Hawking temperature~\cite{Liberati:2020}. Computations of the Hawking temperature always somehow rely on the assumption that the emitted particles move with the speed of light. For instance, the geodesic peeling approach to computing the black hole temperature relies on choosing the null geodesics~\cite{Cropp:2013}. If the gravitons in fact move along timelike/spacelike geodesics, the resulting temperature found in this way will be altered~\cite{Liberati:2020}. That the speed of propagation influences the Hawking effect also becomes apparent in the field of analogue Hawking radiation, where the experimental detection of the effect is possible precisely because the relevant speed is that of sound waves in the medium which is of course much lower than that of light~\cite{Garay:2001,Barcelo:2011,Steinhauer:2016}. Moreover, for certain Horndeski theories of gravity, the first law of black hole mechanics apparently cannot be satisfied without introducing the modified Hawking temperature~\cite{Liberati:2020}. All these theories share an interesting feature. In the Smarr formula~\eqref{smarr}, we have two contributions to the term $-T_{\text{H}}S_{\text{W,inv}}$. The one coming from the integral of the Noether charge over the horizon reads $-\left(\kappa/2\pi\right)S_{\text{W,inv}}$. The second contribution is given by the volume integral of the Noether current and equals $-\left(\kappa/2\pi\right)\left(2\alpha/r_+^2\right)S_{\text{W,inv}}$. Thus, the surface integral over the horizon gives us the standard Hawking temperature $\kappa/2\pi$, whereas the volume integral contributes the correction term to it. The same situation occurs in other Horndeski theories with modified temperature~\cite{Liberati:2020}. At the moment, we cannot say whether this is just a mathematical coincidence or it suggests that the temperature corrections are somehow related to what happens far outside the horizon.

\subsection{Modified speed of gravitons argument}	
	
We can try to evaluate the modified temperature in scalar-tensor Einstein-Gauss-Bonnet gravity directly, using the method proposed for generic Horndeski theories~\cite{Liberati:2020}. The prescription for the modified Hawking temperature in static, spherically symmetric solutions of Horndeski theories depends only on the radial speed of the gravitons~\cite{Liberati:2020}. Metric perturbations in the static spherically symmetric black hole background has been analysed in general Horndeski gravity~\cite{Kobayashi:2012,Kobayashi:2014}. As part of the analysis, an expression for the radial speed of graviton propagation was derived. Unfortunately, metric perturbations of the black hole spacetime we consider diverge on the horizon (due to diverging $\phi'$). Hence, the black hole solution we consider is apparently unstable under perturbations, preventing us from evaluating the corrections to black hole temperature. However, it has been observed that the invariance of the equations of motion under a constant shift of $\phi$ implies that one has a freedom to add a function linear in time to $\phi$ in a static spherically symmetric spacetime. Notably, this has no effect on the metric. Hence, the most general solution time dependent solution for $\phi$ reads
\begin{equation}
\label{phi t}
\phi\left(r,t\right)=\ln\left(\frac{r}{L}\right)\pm\int_{r_+}^{r}\frac{1}{\rho\sqrt{f+L^2\rho^2}}\text{d}\rho+Lt+C,
\end{equation}
where $C$, $L$ are constants. Since function $f$ has a finite minimum $f_{\text{min}}<0$ somewhere between the event and the Cauchy horizon, we have a range of values of $L$ for which $\phi$ is well defined everywhere except at the singularity $r=0$. The additive constant $C$ is then an arbitrary real number. The previously done analysis of perturbations fails for this time dependent $\phi$ (the linear time dependence is only innocuous due to spherical symmetry, which the perturbations break). Therefore, the analysis of both stability and graviton speed in this case needs to be done separately. For these reasons, we are unable to derive our proposed modified temperature~\eqref{T mod} by the modified graviton propagation approach. However, we know that the consistency of the first law of black hole thermodynamics requires this form of the temperature and that the temperature generically appears to be modified in Horndeski theories. Hence, the temperature prescription~\eqref{T mod} seems plausible. Moreover, in the next section, we employ a Euclidean grandcanonical ensemble approach to black hole thermodynamics and show that it leads to the modified temperature~\eqref{T mod}.

\subsection{York-Gibbons-Hawking boundary term}
	
We have seen that adding the Gauss-Bonnet boundary term affects black hole thermodynamics. Since the action should also contain an appropriate York-Gibbons-Hawking boundary term, it is natural to ask whether it influences thermodynamics as well. For completeness, we discuss the boundary terms both for the usually considered action $I_{\text{EGB}}$ and for the shift invariant action $I_{\text{inv}}$.

First, suppose we take the action $I_{\text{EGB}}$ which is not exactly shift invariant and set Dirichlet boundary conditions both for the metric, $\delta g_{\mu\nu}\vert_{\partial\Omega}=0$, and the scalar field, $\delta\phi\vert_{\partial\Omega}=0$. For this case, the boundary term necessary to have a well posed variational problem reads\footnote{As far as we can tell, reference~\cite{Ma:2020} missed the terms proportional to ${^{(3)}G}_{\lambda\rho}K^{\lambda\rho}$ and $n^{\rho}\nabla_{\rho}\phi\nabla_{\lambda}\phi\nabla^{\lambda}\phi$}
\begin{eqnarray}
\nonumber I_{\text{YGH}}&=&\frac{1}{8\pi}\int_{\partial\Omega}\mathcal{K}\sqrt{-h}\text{d}^3x\\
&\equiv&\frac{1}{8\pi}\int\Bigg[K-\frac{2}{3}\alpha\phi\left(K^3-3KK_{\lambda\rho}K^{\lambda\rho}+2K_{\lambda\rho}K^{\rho\sigma}K_{\sigma}^{\;\:\lambda}+3{^{(3)}G}_{\lambda\rho}K^{\lambda\rho}\right) \nonumber\\
&&+2\alpha\left(K^{\lambda\rho}-Kh^{\lambda\rho}\right)\partial_{\lambda}\phi\partial_{\rho}\phi+\frac{2}{3}\alpha n^{\rho}\partial_{\rho}\phi\partial_{\lambda}\phi\partial^{\lambda}\phi\Bigg]\sqrt{-h}\text{d}^3x\,, \label{YGH}
\end{eqnarray}
where $h_{\mu\nu}=g_{\mu\nu}-n_{\mu}n_{\nu}$ denotes the induced metric on the boundary, $n^{\lambda}$ the normal to the boundary (which is taken to be spacelike), ${^{(3)}G}_{\lambda\rho}$ the boundary Einstein tensor, and $K_{\mu\nu}=h^{\lambda}_{\mu}\nabla_{\lambda}n_{\nu}$ its extrinsic curvature. The first term corresponds to the Ricci scalar in the bulk action, the second one to the Gauss-Bonnet term, the third one to the term coupling the Einstein tensor with derivatives of $\phi$ and the last one to the term containing d'Alembertian of $\phi$. The York-Gibbons-Hawking boundary term shifts the Noether charge tensor by $Q_{\text{YGH}}^{\nu\mu}=\frac{1}{4\pi}\mathcal{K}\sqrt{-h}\xi^{[\nu}n^{\mu]}$. If an integral of $Q_{\text{YGH}}^{\nu\mu}$ over the spacetime boundary does not vanish, it will affect its Hamiltonian (and, therefore, the first law of thermodynamics for black hole spacetimes). In our case of static, spherically symmetric black hole spacetime, we have the boundary normal corresponding to a unit, radial vector, i.e., $n^{\mu}=\left(0,\sqrt{f},0,0\right)$. The York-Gibbons-Hawking boundary term then reads
\begin{align}
\mathcal{K}=\frac{f'}{2\sqrt{f}}+2\frac{\sqrt{f}}{r}-8\alpha\phi\frac{f\sqrt{f}}{r^3}+\frac{2}{3}\alpha\phi\frac{f'}{\sqrt{f}}+\frac{2}{3}\alpha\left(\frac{\sqrt{f}-1}{r}\right)^.
\end{align}
Then, we have for the integral of $Q_{\text{YGH}}^{\nu\mu}$ over the horizon
\begin{align}
\nonumber	\int_{\mathcal{H}}Q_{\text{YGH}}^{\nu\mu}\text{d}\mathcal{C}_{\mu\nu}=&\frac{1}{4\pi}\int\left(\frac{f'}{2\sqrt{f}}+2\frac{\sqrt{f}}{r}-8\alpha\phi\frac{f\sqrt{f}}{r^3}+\frac{2}{3}\alpha\phi\frac{f'}{\sqrt{f}}+\frac{2}{3}\alpha\left(\frac{\sqrt{f}-1}{r}\right)^3\right) \\
&fr^2\text{d}\Omega^2\bigg\vert_{r=r_+}=0,
\end{align}
since $f\left(r_+\right)=0$. At the asymptotic infinity, we obtain
\begin{equation}
\int_{\infty}Q_{\text{YGH}}^{\nu\mu}\text{d}\mathcal{C}_{\mu\nu}=\frac{1}{4\pi}\int\left(\frac{f'}{2\sqrt{f}}+2\frac{\sqrt{f}}{r}-8\alpha\phi\frac{f\sqrt{f}}{r^3}+\frac{2}{3}\alpha\phi\frac{f'}{\sqrt{f}}+\frac{2}{3}\alpha\left(\frac{\sqrt{f}-1}{r}\right)^3\right)fr^2\text{d}\Omega^2\bigg\vert_{r\to\infty}.
\end{equation}
This contribution diverges. However, the corrections proportional to $\alpha$ vanish. Hence, the divergent contribution is exactly the same as in general relativity and can be removed by regularisation of the York-Gibbons-Hawking boundary term (i.e., by subtracting $K^0$, the flat spacetime contribution to $K$). The Gibbons-Hawking-York term corresponding to action $I_{\text{EGB}}$ with Dirichlet boundary conditions therefore does not affect black hole thermodynamics in any way.

Let us now consider the York-Gibbons-Hawking term for the exactly shift invariant action $I_{\text{inv}}$. Suppose that we again set Dirichlet boundary conditions, $\delta g_{\mu\nu}=0$, $\delta\phi=0$. The action no longer contains the $\phi\mathcal{G}$ term, and the corresponding boundary term thus no longer appears. It turns out that no extra York-Gibbons-Hawking term is required for the $-\mathcal{G}^{\mu}\nabla_{\mu}\phi$ present in $I_{\text{inv}}$. Thus, we simply have
\begin{equation}
I_{\text{YGH,inv}}=\frac{1}{8\pi}\int_{\partial\Omega}\left[K+2\alpha\left(K^{\lambda\rho}-Kh^{\lambda\rho}\right)\partial_{\lambda}\phi\partial_{\rho}\phi+\frac{2}{3}\alpha n^{\rho}\partial_{\rho}\phi\partial_{\lambda}\phi\partial^{\lambda}\phi\right]\sqrt{-h}\text{d}^3x\,,
\end{equation}
which again does not affect gravitational dynamics. However, there is some tension between the invariance of the action under shifts of $\phi$ by a constant, and fixing the variation of $\phi$ to zero on the  boundary. It would be better to impose a boundary condition on the derivative of $\phi$. A suitable choice is a restricted Neumann boundary condition
\begin{equation}
\label{RNBC}
n^{\nu}\left(\mathcal{G}^{\mu}-8G^{\mu\nu}\nabla_{\nu}\phi-8\nabla^{\nu}\nabla_{\nu}\phi\nabla^{\mu}\phi+8\nabla^{\nu}\phi\nabla_{\nu}\nabla^{\mu}\phi+8\nabla_{\nu}\phi\nabla^{\nu}\phi\nabla^{\mu}\phi\right)\vert_{\partial\Omega}=0.
\end{equation}
If we also impose Dirichlet boundary conditions for the metric and take our action to be $I_{\text{inv}}+I_{\text{YGH,inv}}$, we have a well posed variational problem without the need to fix either $\phi$ or $\delta\phi$ on the boundary. Therefore, the boundary conditions we impose do not break the shift symmetry. Moreover, the boundary condition~\eqref{RNBC} for the derivatives of $\phi$ is automatically satisfied by any solution of the equations of motion, as a consequence of equation~\eqref{G^a}. Hence, we lose no potential solutions by imposing it. Since, as we discussed, the York-Gibbons-Hawking term $I_{\text{YGH,inv}}$ has no influence on black hole thermodynamics, all the conclusions of the previous sections are valid for this choice of boundary conditions.

In summary, the York-Gibbons-Hawking boundary term does not affect the black hole thermodynamics. Therefore, the only boundary term relevant for the definition of the conserved quantities is the Gauss-Bonnet one.

\section{Grandcanonical ensemble derivation of the modified temperature}
\label{Brown-York}

Upon discussing black hole thermodynamics in 4D Einstein-Gauss-Bonnet gravity from the perspective of the covariant phase space formalism, we study the same problem from the perspective of Euclidean canonical ensemble~\cite{BrownYork}. It has the crucial advantage of directly obtaining the black hole temperature by finding stationary points of the Euclidean action. Entropy is then derived by the standard grandcanonical ensemble methods only upon fixing the temperature. Hence, this method supplements the covariant phase space formalism approach, which is unable to directly derive the temperature.
	
In our case, the only variable describing the grandcanonical ensemble is the horizon radius, $r_+$ (other variables, e.g. the electric potential can be easily added~\cite{BrownYork}). The key point of the method lies in introducing an artificial boundary (and the appropriate York-Gibbons-Hawking contribution to action) at finite radial distance. The usual thermodynamic quantities are then recovered in the limit of taking this boundary to infinity.

\subsection{Gauss-Bonnet boundary term in 4D general relativity}
	
As a warm-up and to introduce the machinery of the grandcanonical ensemble method, we show how the Gauss-Bonnet boundary term in 4D changes entropy of a standard Schwarzschild black hole. We work with a general Euclidean ansatz for a spherical, static black hole metric
\begin{equation}
\label{euc metric}
\text{d}s^2=\left(b\left(y\right)\right)^2\text{d}\tau^2+\left(a\left(y\right)\right)^2\text{d}y^2+\left(r\left(y\right)\right)^2\text{d}\Omega_{2}
\end{equation}
where the $\tau$ denotes the Euclidean time coordinates $\tau\in\left[0,2\pi\right)$, and coordinate $y\in\left[0,1\right]$ is chosen so that $r\left(0\right)=r_+$ corresponds to the black hole event horizon and $r\left(1\right)=r_{\text{b}}>r_+$ to the artificial boundary of the spacetime with topology $S^{1}\times S^{2}$. To have the geometry regular at the horizon, we require $b\left(0\right)=0$ and $\left(b'/a\right)_{y=0}=1$~\cite{BrownYork}. The inverse temperature $\beta$ measured by a static observer on the boundary is given by the proper length of its $S^{1}$ component, i.e., $\beta=2\pi b\left(1\right)$~\cite{BrownYork}.
	
The Euclidean action we work with corresponds to vacuum general relativity (for simplicity with $\Lambda=0$) with the Gauss-Bonnet term, i.e.,
\begin{equation}
I=-\frac{1}{16\pi}\int_{\Omega}\left(R+\alpha\nabla_{\mu}\mathcal{G}^{\mu}\right)\sqrt{\mathfrak{g}}\text{d}^4x.
\end{equation}
where $\Omega$ denotes the spacetime. Moreover, we need to fix Dirichlet boundary conditions. Hence, it is necessary to add the appropriate Gibbons-York-Hawking boundary term, which reads (combining the contributions for the Ricci scalar and the Gauss-Bonnet term)
\begin{align}
\nonumber I_{\text{YGH}}=&\frac{1}{8\pi}\int_{\partial\Omega}\left[K-K^0-\frac{2}{3}\alpha\left(K^3-3KK_{\lambda\rho}K^{\lambda\rho}+2K_{\lambda\rho}K^{\rho\sigma}K_{\sigma}^{\;\:\lambda}+3{^{(3)}G}_{\lambda\rho}K^{\lambda\rho}\right)\right]\sqrt{\mathfrak{h}}\text{d}^4x.
\end{align}
In total, the Euclidean action reads
\begin{align}
\nonumber I_{\text{total}}=&-\frac{1}{16\pi}\int_{\Omega}\left(R+\alpha\nabla_{\mu}\mathcal{G}^{\mu}\right)\sqrt{\mathfrak{g}}\text{d}^4x+\frac{1}{8\pi}\int_{\partial\Omega}\bigg[K-K^0 \\
&-\frac{2}{3}\alpha\left(K^3-3KK_{\lambda\rho}K^{\lambda\rho}+2K_{\lambda\rho}K^{\rho\sigma}K_{\sigma}^{\;\:\lambda}+3{^{(3)}G}_{\lambda\rho}K^{\lambda\rho}\right)\bigg]\sqrt{\mathfrak{h}}\text{d}^4x.
\end{align}
	
Evaluating this action for our static spherically symmetric black hole spacetime yields
\begin{align}
\nonumber I_{\text{total}}=&-\int_{0}^{2\pi}\text{d}\tau\int_{0}^{1}\text{d}y\left[-\left(\frac{r^2b'}{2a}\right)'-\frac{ab}{2r'}\left(\frac{rr'^2}{a^2}-r\right)'+2\alpha\left(-\frac{b'}{a}+\frac{r'^2b'}{a^3}\right)'\right] \\
&+\int_{0}^{2\pi}\text{d}\tau\left[-\frac{\left(br^2\right)'}{a}+2br-2\alpha\left(-\frac{b'}{a}+\frac{r'^2b'}{a^3}\right)\right]_{y=1}. \label{action GB}
\end{align}
The constraint equation obtained by varying the symmetry reduced action with respect to $b$ implies $r'/a=\sqrt{1-r_+/r}$, where $r_+$ is an integration constant corresponding to the Schwarzschild radius. Plugging this into equation~\eqref{action GB} and carrying out the integration gives us
\begin{equation}
I_{\text{total}}\left(r_+\right)=\beta r_{\text{b}}\left(1-\sqrt{1-\frac{r_+}{r_{\text{b}}}}\right)-\pi r_+^2-4\pi\alpha,
\end{equation}
where we used $\beta=2\pi b\left(1\right)$. This action corresponds to the grandcanonical partition function for the black hole system we study. To determine the equilibrium configuration, we must find its stationary points with respect to $r_+$, leading to the condition
\begin{equation}
\partial_{r_+}I_{\text{total}}\left(r_+\right)=0,
\end{equation}
which is solved by
\begin{equation}
\beta=4\pi r_+\sqrt{1-\frac{r_+}{r_{\text{b}}}}.
\end{equation}
This corresponds to the redshifted inverse Hawking temperature measured by a stationary observer at a radial finite distance $r_{\text{b}}$ from the black hole. In the limit $r_{\text{b}}\to\infty$ we recover the standard Hawking temperature $T_{\text{H}}=1/4\pi r_+$. The entropy obeys the standard relations
\begin{equation}
S=\beta\left(\frac{\partial I_{\text{total}}}{\partial\beta}\right)_{r_{\text{b}}}-I_{\text{total}}=\pi r_+^2+4\pi\alpha,
\end{equation}
giving the well known constant correction to black hole entropy appearing due to the presence of the Gauss-Bonnet boundary term~\cite{Jacobson:1993,Sarkar:2011}.

\subsection{Modified temperature in 4D scalar-tensor Einstein-Gauss-Bonnet gravity}
\label{mod temp BY}
	
Upon reviewing the approach on the example of general relativity with the Gauss-Bonnet boundary term, we move on to the 4D scalar-tensor Einstein-Gauss-Bonnet gravity. Again considering the metric ansatz~\eqref{euc metric}, we work with the shift symmetric Euclidean action
\begin{align}
\nonumber I_{\text{inv}}=&-\frac{1}{16\pi}\int_{\Omega}\bigg[R-\alpha\mathcal{G}^{\lambda}\nabla_{\lambda}\phi+\alpha\Big(4G^{\lambda\rho}\nabla_{\lambda}\phi\nabla_{\rho}\phi-4\nabla_{\lambda}\phi\nabla^{\lambda}\phi\nabla^{\rho}\nabla_{\rho}\phi \\
\nonumber &+2\nabla_{\lambda}\phi\nabla^{\lambda}\phi\nabla_{\rho}\phi\nabla^{\rho}\phi\Big)\bigg]\sqrt{\mathfrak{g}}\text{d}^{n}x+\frac{1}{8\pi}\int_{\partial\Omega}\Bigg[K-K^0+2\alpha\left(K^{\lambda\rho}-Kh^{\lambda\rho}\right) \\
&\partial_{\lambda}\phi\partial_{\rho}\phi+\frac{2}{3}\alpha n^{\rho}\partial_{\rho}\phi\partial_{\lambda}\phi\partial^{\lambda}\phi\bigg]\sqrt{\mathfrak{h}}\text{d}^3x.
\end{align}
The action already contains the York-Gibbons-Hawking term, which does not include the part corresponding to the Gauss-Bonnet term. This is because, in the shift invariant action, the Gauss-Bonnet term is replaced by $-\alpha\mathcal{G}^{\lambda}\nabla_{\lambda}\phi$ which allows a well-posed Dirichlet boundary conditions for the metric without any York-Gibbons-Hawking boundary contribution. For the scalar field we set the restricted Neumann boundary condition~\eqref{RNBC} which is consistent with the shift symmetry.
	
Evaluating the action for the black hole metric~\eqref{euc metric} yields
\begin{align}
\nonumber I_{\text{EGB}}=&-\int_{0}^{2\pi}\text{d}\tau\int_{0}^{1}\text{d}y\bigg[-\left(\frac{r^2b'}{2a}\right)'-\frac{ab}{2r'}\left(\frac{rr'^2}{a^2}-r\right)'+2\alpha\phi\left(-\frac{b'}{a}+\frac{r'^2b'}{a^3}\right)'+\alpha\frac{\phi'^2}{a^3} \\
\nonumber &\left(-a^2b+2rr'b'+br'^2\right)-\alpha\frac{\phi'^2}{a^2}\left(\frac{br^2}{a}\phi'\right)'+\alpha\frac{br^2\phi'^4}{2a^3}\bigg]+\int_{0}^{2\pi}\text{d}\tau\bigg[-\frac{\left(br^2\right)'}{a}+2br \\
&+2\alpha\phi\left(-\frac{b'}{a}+\frac{r'^2b'}{a^3}\right)+\alpha\frac{br^2\phi'^3}{3a^3}\bigg]_{y=1}.
\end{align}
The constraint equations implied by variation of the action with respect to $b$ requires
\begin{equation}
\zeta\left(r\right)\equiv\frac{r'\left(y\right)}{a\left(y\right)}=\sqrt{1+\frac{r^2\left(y\right)}{2\alpha}\left(1-\sqrt{1+\frac{4\alpha r_+}{r^3\left(y\right)}\left(1+\frac{\alpha}{r_+^2}\right)}\right)}\,,
\end{equation}
where $r_+$ is an integration constant. Moreover, to satisfy the restricted Neumann boundary condition~\eqref{RNBC} for a boundary at an arbitrary $r_{\text{b}}$, we must have
\begin{equation}
\phi\left(y\right)=\ln\frac{r\left(y\right)}{L}-\int_{0}^{y}\frac{a\left(\xi\right)}{r\left(\xi\right)r'\left(\xi\right)}\text{d}\xi.
\end{equation}
where $L$ denotes an integration constant. Plugging this into the action and integrating gives
\begin{align}
I_{\text{inv}}=\beta r_{\text{b}}\left(1-\zeta\left(r_{\text{b}}\right)\right)\left[1-\frac{4\alpha}{3r_{\text{b}}^2}\left(\zeta\left(r_{\text{b}}\right)-1\right)\left(2\zeta\left(r_{\text{b}}\right)-1\right)\right]-\pi r_+^2. \label{euc action}
\end{align}
Looking for stationary points with respect to $r_+$ yields the following condition on $\beta$
\begin{equation}
\label{beta mod}
\beta_{\text{inv}}=4\pi r_+\frac{1}{1-\frac{\alpha}{r_+^2}}\zeta\left(r_{\text{b}}\right)\frac{1-\frac{2\alpha}{r_{\text{b}}^2}\left(\zeta\left(r_{\text{b}}\right)^2-1\right)}{1-\frac{8\alpha}{3r_{\text{b}}^2}\left(\zeta\left(r_{\text{b}}\right)-1\right)\left(3\zeta\left(r_{\text{b}}\right)-2\right)}.
\end{equation}
In the limit of $r_{\text{b}}\to\infty$ we get precisely the modified temperature~\eqref{T mod} we guessed by requiring the validity of the first law of thermodynamics
\begin{equation}
\beta_{\text{inv}}=4\pi r_+\frac{1}{1-\frac{\alpha}{r_+^2}}=\frac{2\pi}{\kappa}\frac{1}{1+\frac{2\alpha}{r_+^2}}.
\end{equation}
Interestingly, at finite distance $r_{\text{b}}$, the inverse temperature~\eqref{beta mod} does not just contain the expected redshift factor $\zeta\left(r_{\text{b}}\right)$, but also an extra correction. We are currently not aware of any natural interpretation of this correction term. For entropy, we obtain
\begin{equation}
S_{\text{inv}}=\beta\left(\frac{\partial I_{\text{inv}}}{\partial \beta}\right)_{r_{\text{b}}}-I_{\text{inv}}=\pi r_+^2,
\end{equation}
i.e., the area law of general relativity.

The Euclidean grandcanonical ensemble approach shows that the modified temperature indeed naturally emerges in 4D scalar-tensor Einstein-Gauss-Bonnet gravity if its shift symmetry is respected by the thermodynamic description. Unfortunately, it provides no clear hints regarding the origin of the temperature modification. Nevertheless, it clearly shows that black hole temperature is sensitive to certain boundary terms appearing in the action. Reasons for this sensitivity require further exploration.

\section{Discussion}
\label{discussion}
	
We have studied black hole thermodynamics in scalar-tensor Einstein-Gauss-Bonnet gravity that is symmetric under the shift of the value of the scalar field by a constant. We argue that one naturally obtains the standard (uncorrected) Bekenstein entropy and a modified Hawking temperature rather than the usual picture of entropy with a logarithmic correction and the standard Hawking temperature. This result is based on three key ideas. First, the shift symmetry ought to be respected by all the physical quantities in the theory, including Wald entropy. Second, the Gauss-Bonnet boundary term affects Wald entropy, even if it does not change the equations of motion. Third, scalar-tensor theories generically have modified propagation speed of gravitons, which may lead to changes in the process of the Hawking radiation (as the majority of the emitted particles are gravitons). This modified temperature can also be derived directly in the Euclidean grandcanonical ensemble approach. While we lack detailed explanation as to how the temperature modifications appear, it might be associated with modified propagation of gravitons in scalar-tensor theories, or perhaps due to a change in thermodynamic ensemble~\cite{Brown:1992bq}. Our reasoning is not limited to the particular action we studied in this work, but applies to any shift-symmetric gravitational theory in four dimensions containing a term of the form $\phi\mathcal{G}$. 
	
It is interesting to ask what impact can our ideas have on thermodynamics of other gravitational theories. For instance, Liouville gravity, a scalar-tensor theory of gravity in two spacetime dimensions is also shift symmetric~\cite{Mann:1994,Frassino:2015}. It depends on the value of $\phi$ only through a term of the form $\phi R$ and the scalar curvature $R$ is a total divergence in two dimensions. Nevertheless, Wald entropy for two dimensional black hole is held to be proportional to the horizon value of $\phi$. Applying the logic we outlined here to this case would instead suggest that the entropy is {\em equal to zero}. This might make some sense given that the horizon spatial cross-section in two dimension is point-like, but the issue certainly deserves a more careful study in the future.
	
More generally, we have shown that certain boundary terms in action both affect Wald entropy and equilibrium temperature of a Euclidean grandcanonical ensemble corresponding to a black hole spacetime. In fact, this is true not just for the shift symmetric theories we discuss above, but essentially for any scalar-tensor theory of gravity. For example, to any scalar-tensor action in four dimensions (among others, any Horndeski theory) we can add a boundary term of the form $\alpha\nabla_{\mu}\left(\phi\mathcal{G}^{\mu}\right)/16\pi$. This leads to a change in the black hole entropy
\begin{align}
\Delta S=\frac{1}{4}\int_{\mathcal{H}}\alpha\phi\left(\epsilon^{\mu\nu}R-4\epsilon^{\lambda\mu} R_{\lambda}^{\;\:\nu}+\epsilon_{\lambda\rho}R^{\nu\mu\lambda\rho}\right)\epsilon_{\mu\nu}\text{d}^2\mathcal{A}.
\end{align}
To then satisfy the first law of black hole mechanics, we must also modify the temperature. In principle, nothing prevents us to add a surface term like this to action, in the same way as we are free to add the Gauss-Bonnet boundary $\alpha\nabla_{\mu}\mathcal{G}^{\mu}/16\pi$ without a scalar field to four dimensional Einstein-Hilbert action. A possibility to affect black hole thermodynamics by boundary terms is not limited to four dimensions. We can similarly add any Lovelock density in its critical dimension, where it becomes a total derivative. Do these terms have a genuine impact on physics that can explain these changes? Or do they signify an ambiguity in both covariant phase space and Euclidean approaches to black hole thermodynamics? In that case, is there a physical way to fix this ambiguity? In the case of 4D scalar-tensor Einstein-Gauss-Bonnet gravity we relied on the shift symmetry as guiding principle, but it is unclear how to approach this problem in general. On the one hand, it might be tempting to fix these ambiguities by demanding the recovery of the standard Hawking temperature. On the other hand, the first law of black hole mechanics then appears to be violated in certain Horndeski theories. Moreover, fixing the standard Hawking temperature is inconsistent with shift symmetry in the case we analysed here. Thus, in our opinion, the possibility of temperature modifications should be taken seriously.
	
Furthermore, the modified propagation speed of gravitons occurs even beyond scalar-tensor theories, e.g. in Lovelock gravity~\cite{Izumi:2014,Papallo:2015,Benakli:2016}. It would be interesting to see whether a modified temperature can occur even in these cases (as speculated in~\cite{MY}). In particular, is it possible that we have entropy in modified gravity given by the Bekenstein formula and move the modifications entirely to the temperature instead? If so, what are the necessary conditions?

A possible way to decide whether the modified thermodynamics we discuss here is physically reasonable would be to check the validity of the second law of thermodynamics. Unfortunately, this is far from straightforward. For example, in the case of four dimensional Einstein-Hilbert action it has been suggested that the Gauss-Bonnet boundary term violates it~\cite{Sarkar:2011}, which could be a reason to exclude it. However, a way to make this term consistent with the second law has been recently proposed~\cite{Gadioux:2023pmw}.

Another possibility is that, in this theory, the first law and Smarr relation receive a new contribution. This contribution could be independent from $T$ and $S$ in the generic case but become degenerate with it in the case of spherical symmetry, leading to the modified temperature. For example, something similar happens for Taub-NUT solutions for the entropy, which in the traditional view receives contributions from the horizon and the Misner string. Testing this idea would be possible once more complicated solutions to the theory have been constructed.

\section*{Acknowledgements}

We would like to thank Pablo Cano for useful comments on our work. 
M.L. is supported by the Charles University Grant Agency project No. GAUK 90123.
RAH is grateful to Andrew Svesko for a number of helpful discussions at the Benasque 2023 workshop, ``Gravity: New Perspectives from Strings and Higher Dimensions''. The work of RAH received the support of a fellowship from ``la Caixa” Foundation (ID 100010434) and from the European Union’s Horizon 2020 research and innovation programme under the Marie Skłodowska-Curie grant agreement No 847648 under fellowship code LCF/BQ/PI21/11830027.
D.K. is grateful for support from GA{\v C}R 23-07457S grant of the Czech Science Foundation.

\appendix

\section{Variational principles in shift symmetric theories: a toy example}

In the main text we have shown that by adding a certain boundary term to the 4D Gauss-Bonnet theory, we can restore its shift symmetry at the level of the action. We also briefly discussed how to then obtain a well-posed variational problem without breaking the shift symmetry in the process. Herein, for the sake of clarity, we discuss this question on a toy example of a 2D gravity. Thus we start from the action:
\begin{equation}
I_0=\int_\Omega d^2x \sqrt{-\mathfrak{g}} \Bigl(\phi R+\frac{1}{2}(\nabla \phi)^2\Bigr)\,,
\end{equation}
where in 2D, we can write $R=\nabla_\mu {\cal R}^\mu$. Such an action is not manifestly shift symmetric, as under $\phi\to \phi+C$ it changes as 
\begin{equation}
\Delta I_0=C\int d^2x \sqrt{-\mathfrak{g}}\nabla_\mu {\cal R}^\mu\,.
\end{equation}
Varying with respect to the scalar field we find 
\begin{eqnarray}
\delta_\phi I_0&=& \int_\Omega 
d^2x \sqrt{-\mathfrak{g}} \Bigl(R\delta \phi +\nabla \phi \cdot \nabla \delta \phi\Bigr)\nonumber\\
&=&
\int_\Omega d^2x \sqrt{-\mathfrak{g}} \Bigl(R-\nabla^2\phi)\delta \phi+\int_{\partial \Omega}dx\sqrt{-\mathfrak{h}}(n\cdot \nabla \phi)\delta \phi\,. 
\end{eqnarray}
We thus have a well posed variational principle with {\em Dirichlet boundary conditions} (DBC) 
\begin{equation}
\delta \phi|_{\partial \Omega}=0\,, 
\end{equation}
which yields the following equation of motion for the scalar field
\begin{equation}\label{EOMapp}
 \nabla^2\phi-R=0\,.
\end{equation}
Alternatively, one can impose a {\em restricted Neumann boundary condition} (RNBC)
\begin{equation}
(n\cdot \nabla \phi)|_{\partial \Omega}=0\,, 
\end{equation}
which of course yields the same EOM. 

The latter can be generalized to the full NBC by adding the corresponding boundary term, namely by considering 
\begin{eqnarray}
I_0^N&=&I_0-\int_{\partial \Omega}dx\sqrt{-\mathfrak{h}}\phi (n\cdot \nabla\phi)\,\nonumber\\
&=&\int d^2x \sqrt{\mathfrak{-g}}\Bigl[(R-\nabla^2\phi)\phi-\frac{1}{2}(\nabla\phi)^2\Bigr]\,,
\end{eqnarray}
which is shift-symmetric on-shell.
The variation then yields 
\begin{equation}
\delta_\phi I_0^N=
\int_\Omega d^2x \sqrt{-\mathfrak{g}} \Bigl(R-\nabla^2\phi)\delta \phi-\int_{\partial \Omega}dx\sqrt{-\mathfrak{h}}\phi \delta (n\cdot \nabla \phi)\,,
\end{equation}
giving the (full) {\em Neumann boundary condition} (NBC):
\begin{equation}
 \delta (n\cdot \nabla \phi)|_{\partial \Omega}=0\,.
\end{equation}

To summarize so far, we have seen that the original action $I_0$ is not shift symmetric. It leads to a well posed variational principle with DBC, which also break the shift symmetry on the boundary. Alternatively, it gives rise to a variational principle with restricted NBC which respects the shift symmetry on the boundary but is not general and will exclude  some physically interesting solutions. On the other hand, the Neumann action $I_0^N$ is shift-symmetric on-shell and the (full) NBC respects the shift symmetry as well. However, to set Dirichlet conditions for the metric, we must add a York-Gibbons-Hawking term of the form $\phi K$ which again breaks the on shell shift symmetry. It has been noted that imposing the Neumann boundary conditions for the metric still allows one to recover the correct black hole entropy~\cite{Khodabakhshi:2020fhb}. However, even this option appears to break the shift symmetry in our setup.  So far, we were unable to find a way to consistently set the boundary conditions, that allows us to both keep the on-shell shift invariance and define the Euclidean canonical ensemble.

Let us finally turn to the type of action studied in the main text:
\begin{equation}
I=I_0-\int d^2x \sqrt{\mathfrak{-g}}\nabla_\mu(\phi {\cal R}^\mu)=
\int d^2x \sqrt{\mathfrak{-g}}\Bigl(-{\cal R}^\mu \nabla_\mu \phi+\frac{1}{2}(\nabla\phi)^2\Bigr)\,.
\end{equation}
This is manifestly shift symmetric even off-shell. Varying this now yields 
\begin{equation}
\delta_\phi I= 
\int_\Omega d^2x \sqrt{-\mathfrak{g}} \Bigl(R-\nabla^2\phi)\delta \phi+\int_{\partial \Omega}dx\sqrt{-\mathfrak{h}}\Bigl[n\cdot (\nabla \phi-{\cal R})\Bigr]\delta \phi\,. 
\end{equation}
This again admits a well posed Dirichlet problem, which, however, breaks the shift symmetry on the boundary. Instead, thus, we can consider the RNBC:
\begin{equation}
(n\cdot A)|_{\partial \Omega}=0\,,\quad A\equiv\nabla \phi-{\cal R}\,.
\end{equation}
Note that in this notation we then also write the EOM, \eqref{EOMapp}, as 
\begin{equation}
\nabla \cdot A=0
\end{equation}
Hence, setting the RNBC does not restrict the solutions of the equations of motion in any way. Moreover,the shift symmetry is respected even off shell.

\section{Alternative boundary term}

A possible problem with using vector $\mathcal{G}^{\mu}$ in the action is that we lack a general expression for it~\cite{Padmanabhan:2011}. However, we can circumvent this issue by considering an explicitly covariant vector
\begin{equation}
\mathcal{S}^{\mu}=8G^{\mu\nu}\nabla_{\nu}\phi-8\nabla^{\nu}\nabla_{\nu}\phi\nabla^{\mu}\phi+8\nabla^{\nu}\phi\nabla_{\nu}\nabla^{\mu}\phi+8\nabla_{\nu}\phi\nabla^{\nu}\phi\nabla^{\mu}\phi.
\end{equation}
By the virtue of equation~\eqref{G^a}, we have $\mathcal{S}^{\mu}=\mathcal{G}^{\mu}$ on shell (we have shown in subsection~\ref{ent section} that the ambiguity in $\mathcal{G}^{\mu}$ has no effect on thermodynamics).

Then, adding boundary term $-\nabla_{\mu}\left(\alpha\phi\mathcal{S}^{\mu}\right)$ to action $I_{\text{EGB}}$~\eqref{action} makes it on shell shift invariant. The expressions for symplectic potential, Noether current and Noether charge obtained from this action read
\begin{align}
\Delta\theta^{\mu}_{\text{EGB}}\left[\delta\right]=&\frac{\alpha}{16\pi}C\delta\mathcal{S}^{\mu}, \\
\Delta j^{\mu}_{\text{EGB},\xi}=&\frac{\alpha}{8\pi}C\nabla_{\nu}\left(\xi^{[\nu}\mathcal{S}^{\mu]}\right) \\
\Delta Q^{\nu\mu}_{\text{EGB},\xi}=&\frac{\alpha}{8\pi}C\xi^{[\nu}\mathcal{S}^{\mu]}.
\end{align}
On shell, these expressions precisely reduce to the ones obtained for the invariant action $I_{\text{inv}}$ we considered in the main text. Since the covariant phase space construction is on shell, we obtain exactly the same conclusions as before, in particular, modified Hawking temperature and Wald entropy proportional to the horizon area.

We may similarly repeat the Brown--York style construction of the Euclidean grandcanonical ensemble. Since $\mathcal{S}^{\mu}$ coincides with $\mathcal{G}^{\mu}$ the final result for the partition function is not going to be affected by swapping both vectors, except we in principle need to add different York-Gibbons-Hawking boundary terms. Unfortunately, we do not know the appropriate York-Gibbons-Hawking term corresponding to $-\nabla_{\mu}\left(\alpha\phi\mathcal{S}^{\mu}\right)$. Nevertheless, we can look for the relevant contributions in the static spherically symmetric case corresponding to the metric~\eqref{euc metric}. There, we only need to  concern ourselves with the terms that do not vanish on the horizon. Since $b\left(r=r_+\right)=0$, these are only the terms proportional to $b'$. For the boundary term corresponding to $-\nabla_{\mu}\left(\alpha\phi\mathcal{S}^{\mu}\right)$, these contribute to the Euclidean action
\begin{equation}
-\left[2\alpha\phi\left(2\frac{b'}{a^3}rr'\phi'-\frac{b'}{a^3}r^2\phi'^2\right)\right]_{r=r_+}=2\alpha\phi
\end{equation}
where we used $b'/a\vert_{r=r_+}=1$, $r'/a\vert_{r=r_+}=0$, $r'\vert_{r=r_+}=1$, and $\phi'=r'/r-a/(rr')$. Compared to the analysis we carried out for $I_{\text{inv}}$ in subsection~\ref{mod temp BY}, we also have to include the York-Gibbons-Hawking boundary term corresponding to $\alpha\phi\mathcal{G}$ which is given in equation~\eqref{YGH}. This term contributes $-2\alpha\phi$ to the Euclidean action, precisely cancelling the term we obtained above. Therefore, we exactly recover the Euclidean action expression~\eqref{euc action}, which implies modified temperature and area law entropy.

To summarize, we have shown that all our results can be reproduced using $\mathcal{S}^\mu$. This has the potential advantage that the covariant expression is known explicitly, as opposed to $\mathcal{G}^\mu$, for which we have an explicit expression only on-shell. Overall, having two paths to the same result strengthens our conclusion.

\bibliography{bibliography}
\bibliographystyle{JHEP}

\end{document}